\documentclass[preprint]{aastex}

\def\gta{\gtrsim}

\def\be { \begin{equation} }
\def\ee { \end{equation} }

\begin{document}


\title{\bf Stellar Pollution in the Solar Neighborhood} 
\author{N.  Murray\altaffilmark{1},
B. Chaboyer\altaffilmark{2}, P. Arras\altaffilmark{1}, 
B. Hansen\altaffilmark{1,3} and R. W. Noyes\altaffilmark{4}}

\altaffiltext{1}{Canadian Institute for Theoretical Astrophysics,
 60 St. George st., University of Toronto, Toronto, ONT M5S 3H8,
Canada; murray, arras@cita.utoronto.ca}

\altaffiltext{2}{Department of Physics and  Astronomy , Dartmouth
College, 6127 Wilder Laboratory, Hanover, NH 03755-3528, USA;
chaboyer@heather.dartmouth.edu }     

\altaffiltext{3}{present address: Department of Astrophysical
Sciences, Princeton University, Princeton, NJ 08544-1001 USA;
hansen@astro.princeton.edu}

\altaffiltext{4}{Harvard-Smithsonian Center for Astrophysics, 60
Garden Street, Cambridge, MA, 02138, USA; noyes@cfa.harvard.edu}

\begin{abstract}
We study spectroscopically determined iron abundances of $642$
solar-type stars to search for the signature of accreted iron-rich
material. We find that the metallicity [Fe/H] of a subset of $466$ main sequence
stars, when plotted as a function of stellar mass, mimics the pattern
seen in lithium abundances in open clusters. Using Monte Carlo models
we find that, on average, these stars have accreted $\sim0.4M_\oplus$
of iron while on the main sequence. A much smaller sample of $19$
stars in the Hertzsprung gap, which are slightly evolved and whose
convection zones are significantly more massive, have lower average
[Fe/H], and their metallicity shows no clear variation with stellar
mass. These findings suggest that terrestrial-type material is common
around solar type stars.
\end{abstract}


\section{INTRODUCTION}

The discovery that at least $6-8\%$ of solar type stars harbor
Jupiter-mass or larger bodies, often in small, eccentric orbits \citep{mq,
marcy, butler, mcm} shows that
planets are not exceptionally rare. The transiting planet orbiting
HD 209458 has a mass of $0.69$ Jupiter masses, and a radius
about 1.4 times that of Jupiter. Clearly it is a gas giant like
Jupiter or Saturn; the minimum masses of the other known objects
suggest that they are also gas giants. The Doppler technique
used for most of the discoveries cannot find terrestrial
mass objects in AU scale orbits. What fraction of solar type stars
have terrestrial-type planets?

How could an observer with our current technology located on a star in
the solar neighborhood decide if there were any terrestrial-type
bodies orbiting the sun?  More generally, how could such an observer
estimate the fraction of solar type stars having companions of
terrestrial (as opposed to gas giant) compositions? In this paper we
explore one possible way of addressing the latter question,
stellar-mass dependent photospheric metallicity trends.

In section \ref{pollution} we argue that a few Earth masses of
rocky/icy material have accreted onto the Sun over its lifetime. In
section \ref{evidence} we discuss one possible observational tracer of
similar accretion occurring on other solar type stars. In section
\ref{observations} we examine a sample of stars with spectroscopically
measured photospheric iron abundances 
\be 
{\rm [Fe/H]}\equiv\log\left[{f_{Fe}\over f_{Fe,\odot}}\right],
\ee 
where $f_{Fe}$ is the mass abundance of iron in the photosphere of the
star, and $f_{Fe,\odot}\approx 1.3\times10^{-3}$ is the the mass abundance
of iron in the photosphere of the sun.  We compare the observations to
Monte Carlo models of stellar pollution in section \ref{gamble}. We
discuss the implications of our results in section \ref{discussion}
and present our conclusion in the final section.

\section{POLLUTION BY SMALL BODIES IN OUR SOLAR SYSTEM
\label{pollution}}

\begin{figure}
\plotone{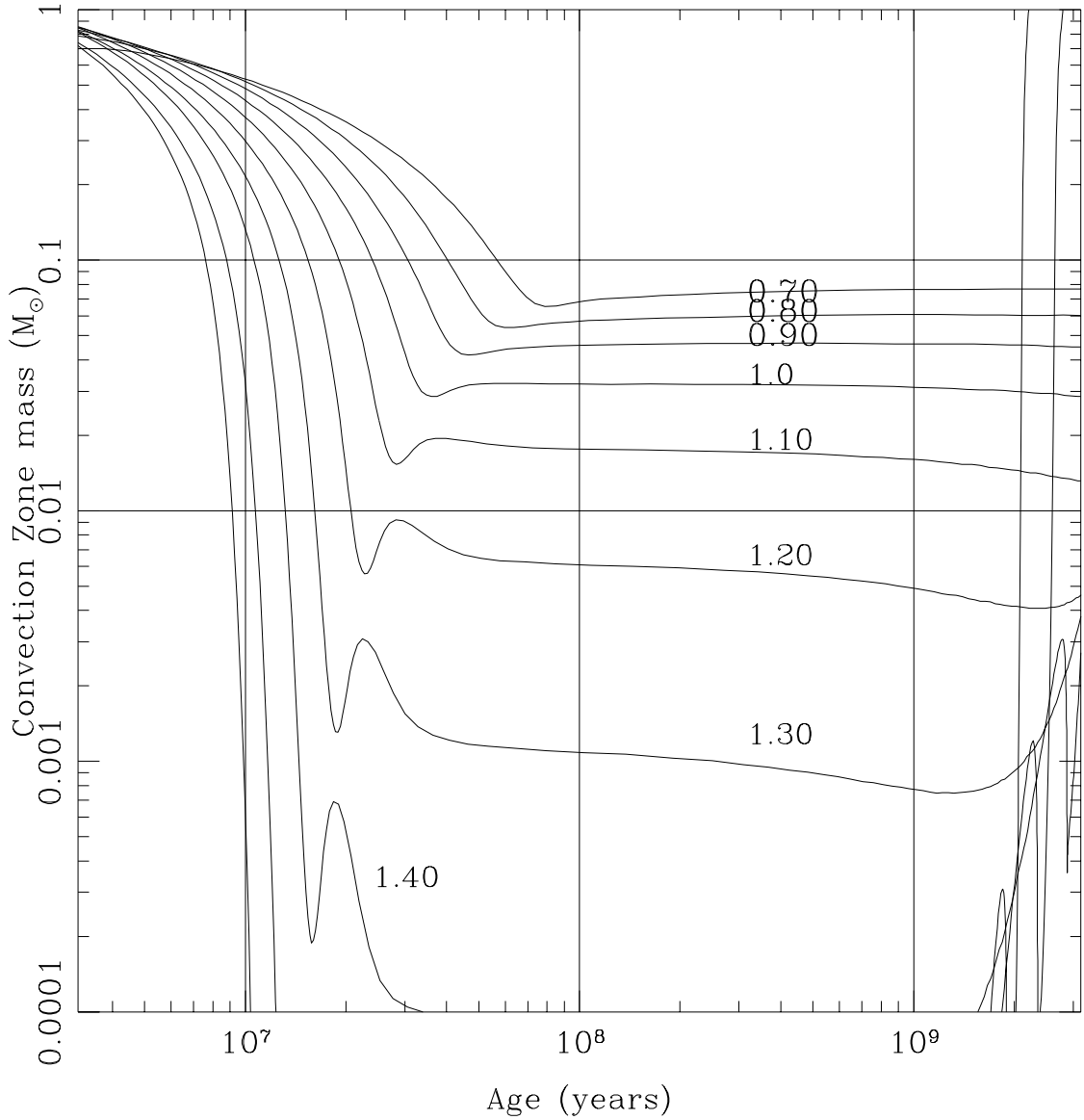}
\caption[Convection Zone Mass]{The mass of the convection zone for
solar metallicity (${\rm [Fe/H]}=0$) stars, plotted as a function of stellar age. The stellar
mass is used to label the curves. Note that for ages less than about
7 million years all stars in the mass range $0.6-1.6M_\odot$ have
convection zone masses greater than $0.1M_\odot$;  stars this young have not
yet reached the main sequence. 
\label{Fig_conv_mass}}
\end{figure}

The surface density $\Sigma_{Fe}$ of iron in our solar system follows
a rough power law with distance from the sun
\cite{weidenschilling}. We have compiled recent estimates for the
iron content of the planets in table~\ref{Table_iron}. From a least
squares fit including Venus, Earth, and the four gas giants we find
$\Sigma_{Fe}\sim 4(r/1\, AU)^{-1.7}{\rm \,g\,cm^{-2}}$.
Weidenschilling noted a clear deficit of material in the region
interior to Venus, and between Earth and Jupiter, relative to this
power law. Gas drag acting on planetesimals near the sun could reduce
the surface density of such bodies near the present orbit of Mercury
by dropping them onto the sun. Because the solar convection zone was
very deep for $\sim20$ million years (see figure~\ref{Fig_conv_mass}),
while the gas disk that would have produced the drag is believed to
have survived only a few million, this would not have altered the
apparent metallicity of the sun.

Materiel in the asteroid belt clearly survived such gas drag. However,
Kirkwood noticed over one hundred years ago that the distribution of
asteroids showed distinct gaps at the location of orbital resonances
with Jupiter \cite{kirkwood}. This suggests that material has been
removed from the gaps under the influence of Jupiter. Over the last
twenty years, the dynamics of this removal have been worked out in
considerable detail \cite{Wisdom, hm, Gladman}. The gaps
are the result of resonant, chaotic perturbations of the asteroid
orbits by Jupiter. A second feature, seen in plots of orbital
eccentricity versus semimajor axis, is the result of a secular
resonance, the $\nu_6$ resonance. In either type of resonance bodies
with semimajor axis larger than about half that of Jupiter's tend to
be removed from the solar system; their eccentricities grow with time,
increasing their apoapses until they reach the orbit of Jupiter. Close
encounters with that planet then quickly eject the asteroids.

Resonant asteroids in smaller orbits tend not to be ejected; their
apoapse does not reach the orbit of Jupiter. Instead, their periapses
decrease until they hit the sun.  Several related mechanisms can
combine to produce this decrease in periapse, but the two main actors
are mean motion resonances and the $\nu_6$ resonance already
mentioned.  In a mean motion resonance such as that responsible for
the $3/1$ Kirkwood gap, the eccentricity of the resonant asteroid
undergoes a random walk. Since the eccentricity cannot decrease below
zero, there is a tendency for bodies with small $e$ to drift to larger
$e$. When the eccentricity is large enough, the asteroid strikes the
sun \cite{Gladman}. A small fraction will suffer close encounters with
the Earth or other terrestrial bodies; this may lead to Jupiter
crossing orbits, which usually leads to ejection of the asteroid from
the solar system. Only a very small fraction will collide with either a
terrestrial planet or Jupiter. In the $\nu_6$ secular resonance the
precession rate equals the sixth secular frequency of the solar
system. This frequency appears with substantial amplitude in the
orbital motion of both Jupiter and Saturn; it is roughly the average
precession rate of the latter planet. The $\nu_6$ resonance rapidly
removes angular momentum from the orbit of the asteroid, dropping the
periapse below the surface of the sun. 

These resonances are currently active, and are believed to be
responsible for the bulk of the meteorites striking the Earth. While
there is no direct observational confirmation, it is virtually certain
that a much larger flux of asteroidal material from these resonances
is currently hitting the sun. 

It is also likely that the flux of asteroidal material was much larger
in the past.  The locations of both mean motion and secular resonances
depends on the positions of Jupiter and Saturn, and on any
massive gas disk. It has been suggested that as the protoplanetary gas
disk dissipated the $\nu_6$ resonance swept through the belt,
depleting much of the mass. More recently it has been suggested that
one or both of Jupiter and Saturn migrated to their current positions,
causing a similar sweeping of mean motion resonances. This would
preferentially deplete the outer belt \cite{hm, liou}, but
would also drop material from the inner belt onto the sun.  Finally,
we note that the asteroid belt is hot, in the sense that the
distribution of both $e$ and $i$ is broad. A number of authors have
suggested that a planet size body swept through the region early in
the history of the solar system, heating the belt as well as dropping
material on the star \cite{wetherill, pmv}.

The surfaces of the terrestrial planets and the Moon show evidence for
an extended period of bombardment, known as the late heavy
bombardment, lasting up to a billion years after their surfaces
cooled. This is consistent with some of the longer lived depletion
scenarios described above.

From table~\ref{Table_iron}, the amount of material inferred to have
populated the original asteroid belt is of order five to ten Earth
masses. Roughly 2-5 Earth masses would have been between the present
position of Mars and $2.5$ AU, half the present semimajor axis of
Jupiter. A substantial fraction of this material, of order half, would
have ended up in the sun. We will take $\sim2M_\oplus$ as a
representative number. Meteoritic material is roughly $20\%$ iron by
weight \cite{GA}, so we take $\sim0.5M_\oplus$ as the amount of iron
that was dropped on the sun after the convection zone thinned.

\section{POSSIBLE OBSERVATIONAL EVIDENCE FOR POLLUTION\label{evidence}}

There is another possible piece of evidence besides the late heavy
bombardment for such a late depletion
of the asteroid belt. As we noted above, a substantial fraction of the
``missing'' material originally in the asteroid belt strikes the
sun. This material will be mixed throughout the convection zone of the
star. Since a substantial bombardment of the terrestrial
planets lasted for much longer than 100 million years, this material
would have landed in a convection zone of mass
$\sim3\times10^{-2}M_\odot$. Hence there is a possibility that the
signature of  solar bombardment was left in the form of an enhanced
metallicity in the solar envelope.

The mass of the convection zone was $\sim3\%$ of the
solar mass between $10^8$ and $10^9$ years, while the present mass
fraction of iron observed in the photosphere is
$1.3\times10^{-3}$. This yields a total iron content of
$\sim12.75M_\oplus$ between $10^8$ and $10^9$ years. Adding half an
Earth mass of iron would increase the observed ${\rm [Fe/H]}$ by only
$\approx 0.017$ dex.

Precision measurements of the bulk metallicity of a single solar type
star are not presently available. Measurements of the sound speed in
the interior of the sun, using the five minute oscillations, could
in principle provide such a measurement. The only work along these lines
that we are aware of is the paper by Henney \& Ulrich (1998). They
show that the addition of $\sim8M_\oplus$ of cometary material results in a
frequency shift that, because of uncertainties in solar models, is too
small to be detected.

However, it is possible that observations of large numbers of stars
could reveal the presence of pollution. For example, stars having a
fixed metallicity but different masses will mix material dropped onto
their surfaces to different depths. Knowledge of the depth of this
surface mixing layer, which we summarize in the next subsection,
allows us to look for variations of metallicity with stellar mass
produced by accreted material.

\subsection{The Mass Of The Surface Mixing Region}\label{Section_mixing} 

Material dropped onto the surface of a star will not in general be
confined to the photosphere. The surface layers of stars less than
about $1.4M_\odot$ are convectively unstable, and any material dropped
on the star is expected be mixed throughout the convection
zone. Observations of lithium (Li) abundance provide support for this
theoretical picture. But these same observations 
suggest that there is another form of mixing in stars more massive than
$\sim1.2M_\odot$, possibly related to meridional circulation.

The evidence for the two types of mixing comes in the form of
variations in the surface lithium abundance with stellar mass. Lithium
is destroyed when it is exposed to temperatures above about
$3.1\times10^6K$; the temperature at the photosphere is much lower
than this, so lithium depletion indicates that photospheric material
is mixed down to regions where the temperature is higher. Lithium
could also be removed from the photosphere by simple settling (so that
the lithium is stored below the photosphere, as opposed to being
destroyed) or by stellar mass loss. However, the observations, which
we review below, point to destruction by mixing.

Observations show that for stars with mass below $\sim1.2M_\odot$ the
photospheric abundance of lithium decreases rapidly with decreasing
stellar mass,  e.g., \cite{boesgaard}. Some depletion occurs
before the main sequence, but there is strong evidence that depletion
also occurs on the main sequence \cite{jfs}. The
main sequence depletion is believed to be related to the increasing
depth of the bottom of the surface convection zone with decreasing
stellar mass in these stars, supplemented by some convective overshoot
and/or settling in stars near the upper end of this mass range.

Stars above about $1.2M_\odot$ have very thin, or even lack,
convection zones.  In standard stellar models, with no settling, mixing or
mass loss, the surface Li and Be abundances are constant over the main
sequence lifetime of such high mass stars.  Both Li and Be are destroyed below
modest depths in the interiors of such stars. For stars in the mass
range $1.2$-$1.6M_\odot$ and having solar metallicity, our stellar
models indicate that the nominally undepleted region has a mass of
$\gta3\times10^{-2}M_\odot$ for Li and $\gta6\times10^{-2}M_\odot$ for
Be. (We assume that the lithium is destroyed at temperatures above
$3.1\times10^6$K). At depths below those corresponding to these masses,
Li and Be should be destroyed.

Real stars with masses larger than $1.2M_\odot$ don't behave like the
models. Hyades stars show a distinct ``lithium dip'' centered around
$1.4M_\odot$ \cite{bt86}. Between $\sim 1.2$ and $1.4M_\odot$ the
lithium abundance is seen to decrease with increasing stellar
mass. From $1.4$ to $1.5M_\odot$ there is a sharp increase in lithium
abundance.  Balachandran (1995) finds the highest abundances for stars
with masses above $1.5M_\odot$ (the ``blue side'' of the dip) and for
stars of $\approx1.2M_\odot$ (the ``red side'' of the dip); these
stars have abundances about equal to those found in meteorites
\cite{GA}, abundances which are generally believed to be primordial.

Observations of young stellar clusters such as $\alpha$ Per and the
Pleiades find little evidence of lithium depletion in more massive stars
\cite{soderblom}. Indeed, there is a strong correlation between
cluster age and Li abundance in the ``lithium dip'' between $1.2$ and
$1.5M_\odot$, indicating that the depletion takes place on the main
sequence on a time scale of 100 Myrs or more \cite{boesgaard}.

Observations of clusters having different metallicities ${\rm [Fe/H]}$ show
that the location of the dip is slightly metallicity
dependent \cite{balachandran}. Lower metallicity clusters have dips
centered at lower masses. However, the high mass side of
the dip depends only weakly on metallicity.

As we noted above, the lithium depletion appears to be due to mixing rather than
gravitational settling or stellar mass loss. Two observations lead to
this conclusion. First, surface depletions of lithium and beryllium
are correlated, with the surface lithium diminishing more rapidly than
surface beryllium \cite{deliyannis}. Because the Be rich zone has
twice the mass of the Li rich zone, mass loss would deplete all the Li
before depleting the surface Be. Similarly, settling predicts that
surface Li and Be would be depleted at roughly the same rate.  Second,
evolved stars with $1.2<M<1.5$ are also deficient in lithium
\cite{gilroy, balachandran}; settling of lithium would leave
a non-depleted layer of lithium below the surface (in the absence of
mixing), which would be dredged up to the surface when the convection
zone deepens. The fact that the predicted increase in evolved stars is
not seen indicates that the Li missing from main sequence stellar
photospheres is destroyed rather than sequestered below the
photosphere.

Above $1.5M_\odot$ there appears to be little mixing of the upper
layers of the star with deeper layers, since there is little Li
depletion. However, the lithium in these stars does appear to be
destroyed in deeper layers; evolved (giant) 1.6 solar mass stars are
severely depleted \cite{gilroy}. This depletion occurs inside the
progenitors, below the photosphere \cite{vauclair}, so that when the
surface convection zone forms and then thickens, the surface lithium
is mixed over a larger volume of depleted material, reducing the
photospheric lithium abundance.

The surface mixing seen in the dip appears to be rotationally induced
\cite{charbonnel, balachandran}, 
although this is far from certain.

As already noted, stars in $\alpha$ Per and the Pleiades clusters,
with estimated ages of 50 and 70 million years, show little depletion
in Lithium abundance for stars between $1.2$ and $1.5$ solar masses,
while stars in the Hyades and older clusters show depletions that
increase with cluster age. We interpret this to mean that the upper
layers of the stars are mixed downward on time scales of $\sim 100$
million years.

\begin{figure}
\plotone{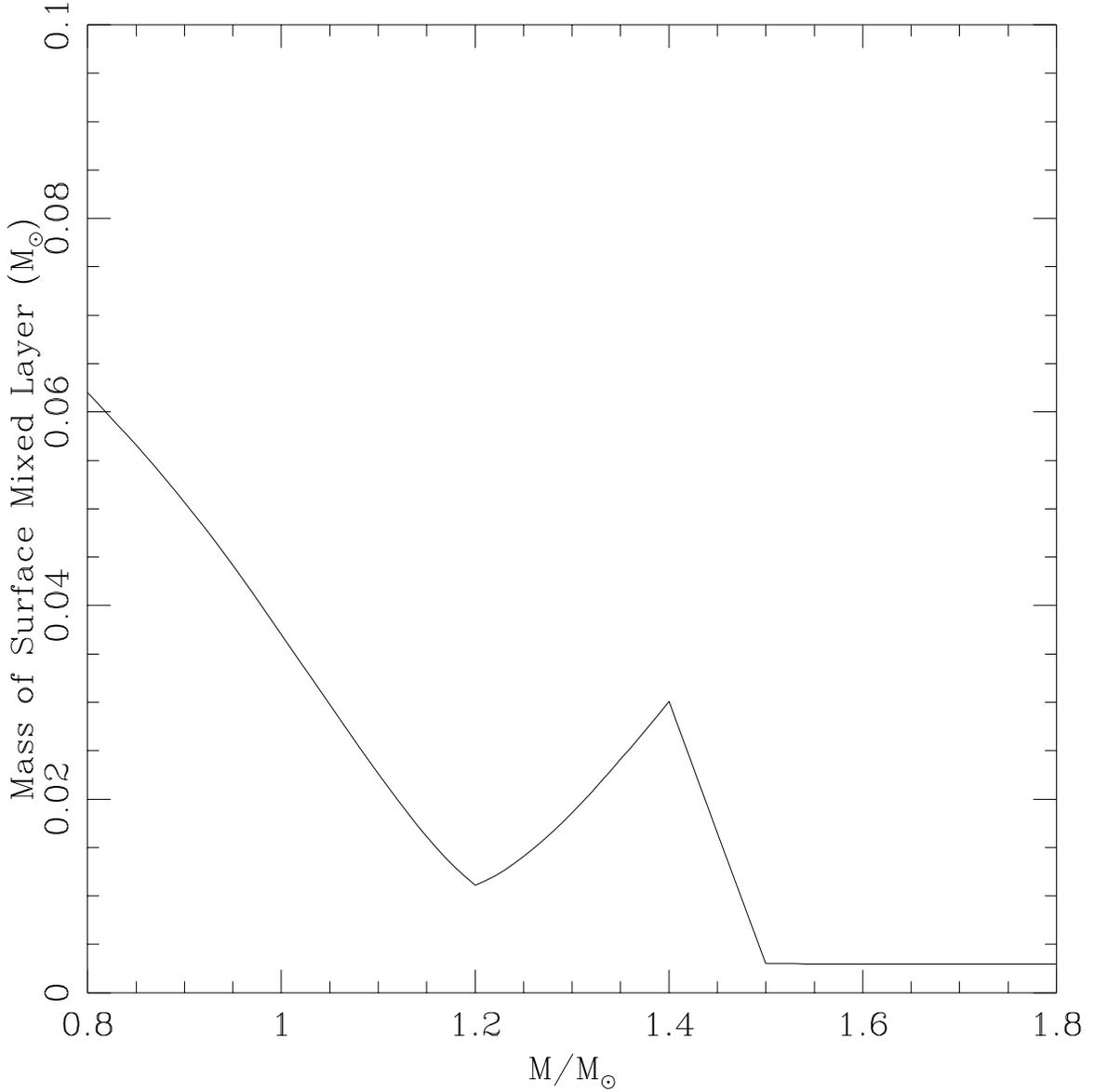}
\caption[Mass of Surface Mixing Layer]{The mass of the surface mixing
layer of solar metallicity stars. For masses below about $1.0M_\odot$
the mass is assumed to be the mass of the convection zone, as
calculated in our stellar models. Above $1.5M_\odot$ the mass is
assumed to be $3\times10^{-3}M_\odot$, as predicted by  meridional
circulation models. Between these masses the mixed region is
chosen by comparison to observations of the lithium dip. See the text
for a more detailed explanation. 
\label{Fig_mixing_mass}}
\end{figure}

The observational data summarized here leads us to propose the
following simple model for the mass $M_{mix}$ of the surface mixing
region in solar mass stars (see Figure \ref{Fig_mixing_mass}). We
first describe the lithium dip. The location, in mass, of the dip is
described by $M_{red}$, $M_{central}$, and $M_{blue}$, giving the low,
central, and high mass points of the dip. These three masses depend on
the stellar ${\rm [Fe/H]}$.

 Above $M_{blue}$ we assume $M_{mix}=M_{mh}$; we
use $M_{mh} =3\times10^{-3}M_\odot$, the value obtained for the
meridonal circulation model of Charbonnel et al (1992). This number is
currently unconstrained by observation on the low side, but must be
less than $\approx5\times10^{-3}M_\odot$ to prevent Li depletion. 

For $M_{central}\le M_*<M_{blue}$ we use a linear interpolation
between $M_{mh}$ and $M_{mc}\approx3\times10^{-2}$, where the latter
value (for a star at the center of the dip) allows for complete Li
depletion but only partial Be depletion. We also use a linear
interpolation between $M_{mc}$ and $M_{ml}$ for $M_{red}\le
M_*<M_{central}$. Stars with $M=M_{red}$ appear to be only slightly
depleted, so we take $M_{ml}\approx 5\times10^{-3}M_\odot$.

Extrapolating from data in \cite{balachandran}, we take
\begin{eqnarray}  
M_{red}/M_\odot=&1.51+0.054{\rm [Fe/H]}\\
M_{central}/M_\odot=&1.42+0.25{\rm [Fe/H]}\\
M_{blue}/M_\odot=&1.31+0.31{\rm [Fe/H]}.
\end{eqnarray} 
Note that the high mass or blue side of the dip is essentially
independent of ${\rm [Fe/H]}$.
For stars with mass below $M_{red}$, we take the mass of the
convection zone ($M_{cvz}$) plus the mass of the mixing zone $M_{ml}$
of a $M_{red}$ star on the low mass or red edge of the dip.

The mass of the convection zone at an age of $10^8$ years (the mass at
$10^9$ is similar, see Figure \ref{Fig_conv_mass}) is found from the
following empirical fit, derived from our stellar models:
\be \label{eqn_cvz_mass} 
M_{cvz}=a\left({\rm [Fe/H]}\right)\times(M_{-3}-M_*)^{3.65},
\ee 
where $a\left({\rm [Fe/H]}\right)=-0.2{\rm [Fe/H]}+0.49$. The quantity $M_{-3}$ is the
(metallicity dependent) stellar mass at which the convection zone mass
equals $10^{-3}M_\odot$,
\be 
M_{-3}=a_{-3}{\rm [Fe/H]}+b_{-3},
\ee 
where $a_{-3}=0.32$ and $b_{-3}=1.45$. Equation (\ref{eqn_cvz_mass})
is very accurate near $M_{-3}$ and good to about $20\%$ at
$0.8M_\odot$.

Our model for the depth of the mixing region in a solar metallicity
star is shown in Figure~(\ref{Fig_mixing_mass}). Note that stars of
different metallicity will have ``lithium dips'' (peaks in this
figure) at slightly different masses than those shown in the figure.

Armed with this model, we can predict the average change in
metallicity of a sample of stars subjected to iron
accretion. Essentially we will look for behavior similar to that seen
in the abundance of lithium. Polluted low mass stars of a fixed age
will show an increase in metallicity with increasing mass (below the
low mass end of the Li dip, around 1.0 solar masses). Intermediate
mass polluted stars will show a decrease in [Fe/H] with increasing
mass, up to $M_{dip}$, followed by a sharp increase at $M_{blue}$.

We want to stress that, while we are using the lithium data as a proxy
to estimate the depth of the surface mixing layer of main sequence
solar type stars, and while we anticipate that the surface abundances
of lithium and iron will follow similar patterns, the reason for the
abundance variations differ. The lithium abundances vary because
lithimum is destroyed when it is mixed below the surface of the
star. Stars in the mass range we consider neither destroy nor create
iron. Rather, we assume that some iron is added to the surface of the
star after it forms; it is then diluted by the mixing process down to
the same depth that the lithium is mixed to. In other words the
variations in lithium are produced by the destruction of lithium
inside the star, while the variations in iron are produced by the
addition of iron from outside the star.

\section{OBSERVATIONAL CONSTRAINTS ON POLLUTION\label{observations}} 

In order to look for mass-dependent variations in [Fe/H], we must be
able to estimate the masses of stars, and we must have reliable
measurements of the metallicity. To find the mass of a star we need to
know its luminosity, or alternately its absolute V magnitude $M_v$,
its colors (we use $B-V$), and its composition [Fe/H]. The
HIPPARCOS catalogue gives parallaxes accurate to roughly one
milliarcsecond for a selection of stars in the solar neighborhood,
allowing us to find absolute magnitudes when we are given $V$. We use
SIMBAD values for $B$ and $V$. We take metallicities from the
\citep{Cayrel} catalogue. This lists spectroscopically determined
values for [Fe/H] taken from the literature.

We take all the HD stars in the Cayrel de Strobel et al. catalogue for
which there are HIPPARCOS parallaxes larger than 10 milliarcseconds
(corresponding to distances less than 100pc). We also require that the
error in the parallax be less than 10 percent. Since we are interested
in main sequence stars that have masses less than about $2M_\odot$, we
eliminate all stars with $M_v<1.0$. We then fit the remaining stars to
stellar tracks taken from a grid of models having $-0.5\le {\rm [Fe/H]}\le
0.5$ in steps of $0.05$ dex, and masses in the range $0.60\le M\le
1.75M_\odot$ in steps of $0.05M_\odot$. The models are from the Yale
stellar evolution code.

A star is defined to be evolved if it has a convection zone mass more
than ten times larger than the mass of the surface mixing layer at an
age of $10^8$ years, while the absolute $V$ magnitude has not
decreased by more than $\approx0.5$.  We refer to such stars as
``Hertzsprung gap'' stars.  This occurs roughly when the star leaves
the main sequence.  It is a useful definition because it allows us
another check on the pollution scenario; if pollution is occurring,
the class of stars with convection zone masses ten times larger than
their initial main sequence convection zone masses should show
substantially lower [Fe/H] values than their parent populations.  

If the absolute $V$ magnitude has decreased by more than $\approx0.5$,
the star is considered to be a giant. Because their surface gravities
are so much lower than main sequence stars of the same mass, we
hesitate to compare the metallicity trends of giants and dwarfs.

In the course of our analysis a problem with our stellar models became
apparent. Stars with $B-V>1$ were systematically found to be older
than 20Gyrs. We believe that this is due to a failure of our
atmospheric models at these low effective temperature. For this reason
we do not include any stars redder than this in our analysis; this
corresponds to stars with masses less than $\sim0.75-0.80M_\odot$. At
the other end of the mass spectrum, we felt confident in extrapolating
up to masses of $1.80M_\odot$, $0.05M_\odot$ beyond our grid.

Of the 650 stars in our sample, 466 were unevolved and had masses in
the range $0.8\le M\le1.8$. Fifteen stars had convection zones 3-10
times larger than their main sequence value (at $10^8$yrs); we treated
these as unevolved. Nineteen stars had convection zones more than ten
times as massive as the main sequence values; these are our Hertsprung
gap stars.  

\begin{figure}
\plotone{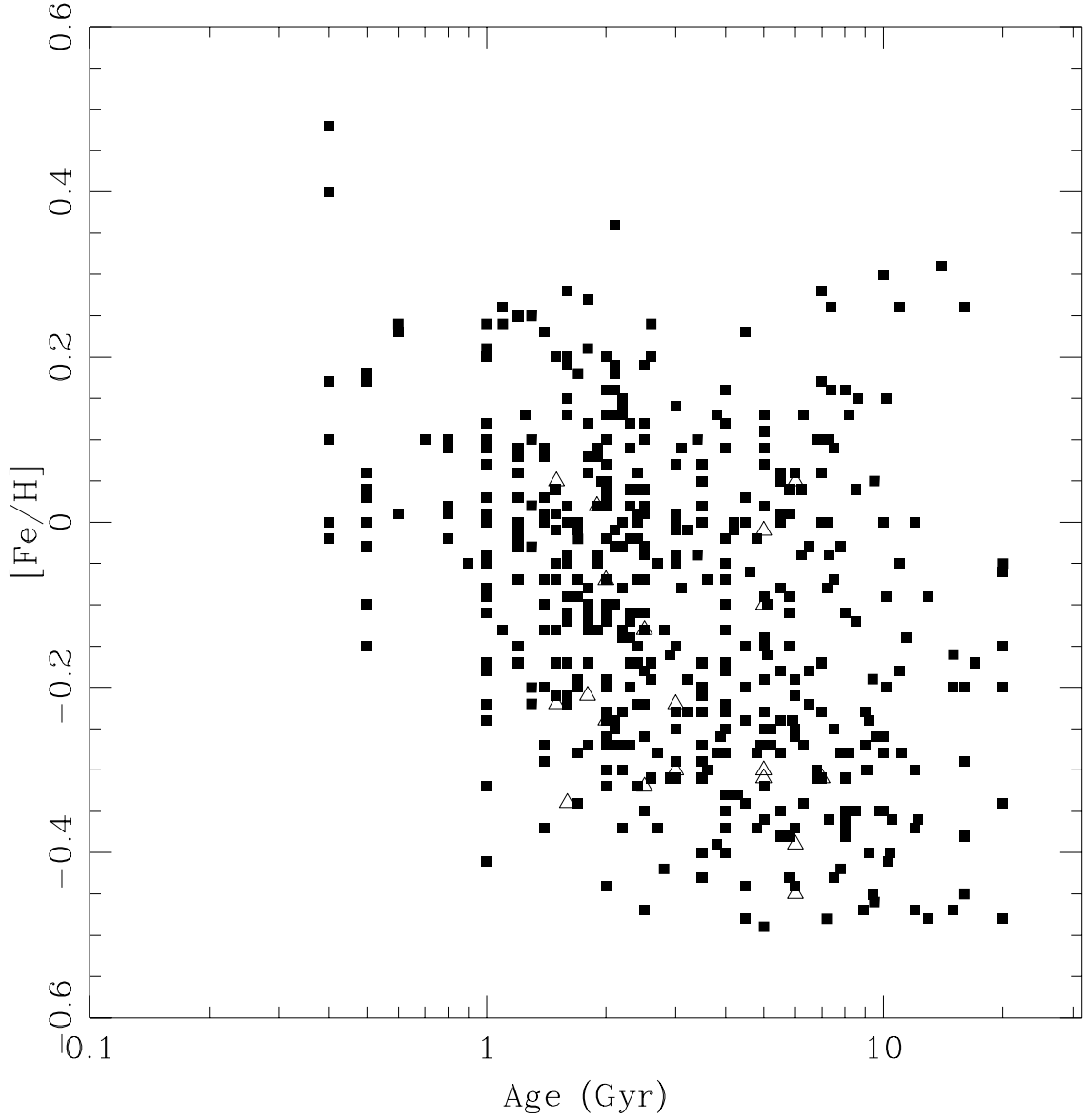}
\caption[Metallicity versus age]{Stellar metallicity as a function of
the logarithm of the stellar age, where the latter is obtained by
fitting to our stellar models. Filled squares represent unevolved
stars, while open triangles represent  evolved stars (dwarf
stars with surface convection zones ten or more times larger than
$M_{mix}$. 
\label{Fig_age}}
\end{figure}

\begin{figure}
\plotone{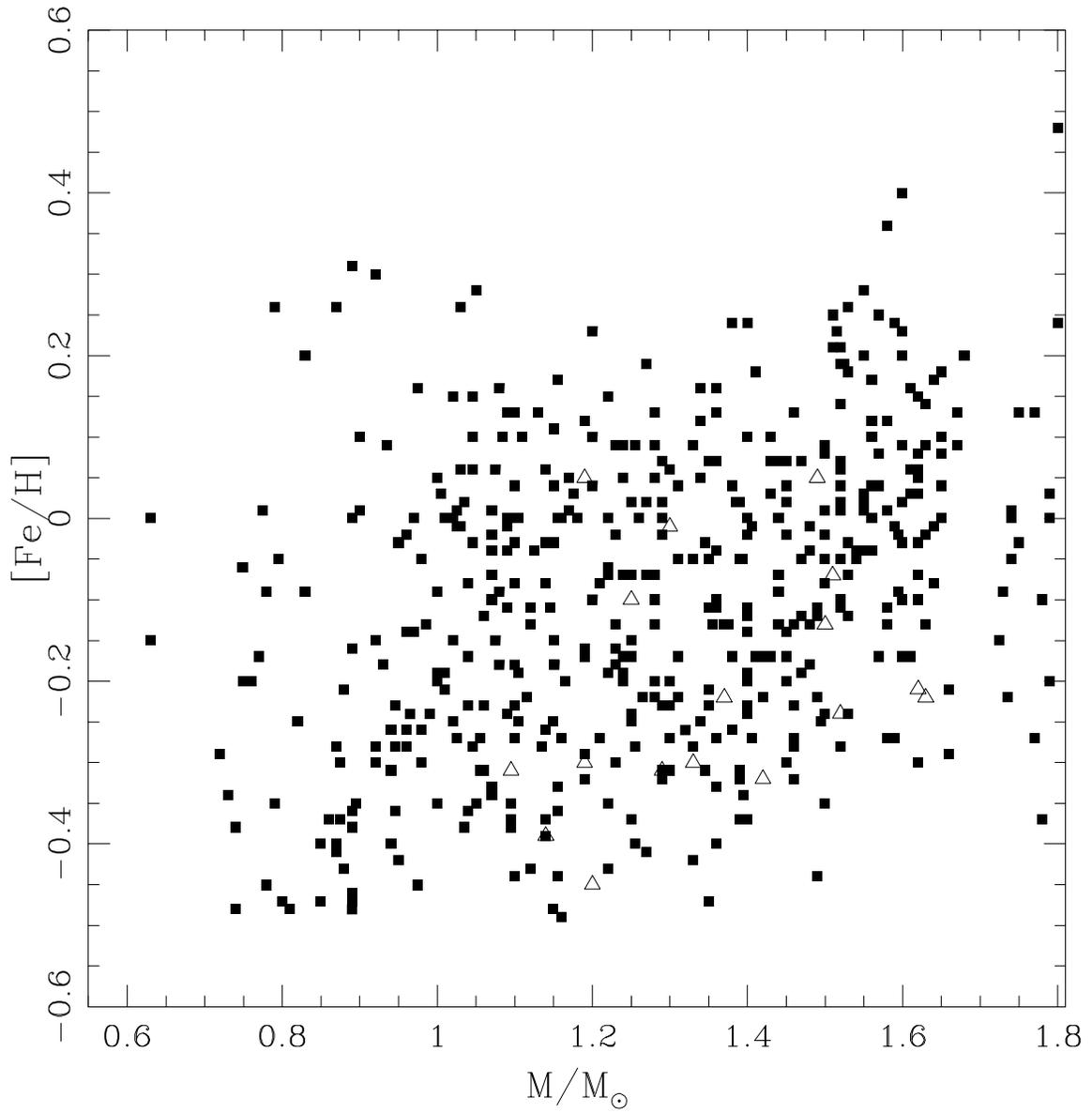}
\caption[Metallicity versus stellar mass]{Stellar metallicity as a
function of stellar mass, where the mass is obtained by fitting to our
stellar models. As in Figure \ref{Fig_age}, filled squares represent unevolved
stars, while open triangles represent evolved stars.  
\label{Fig_mass}}
\end{figure}

\begin{figure}
\plotone{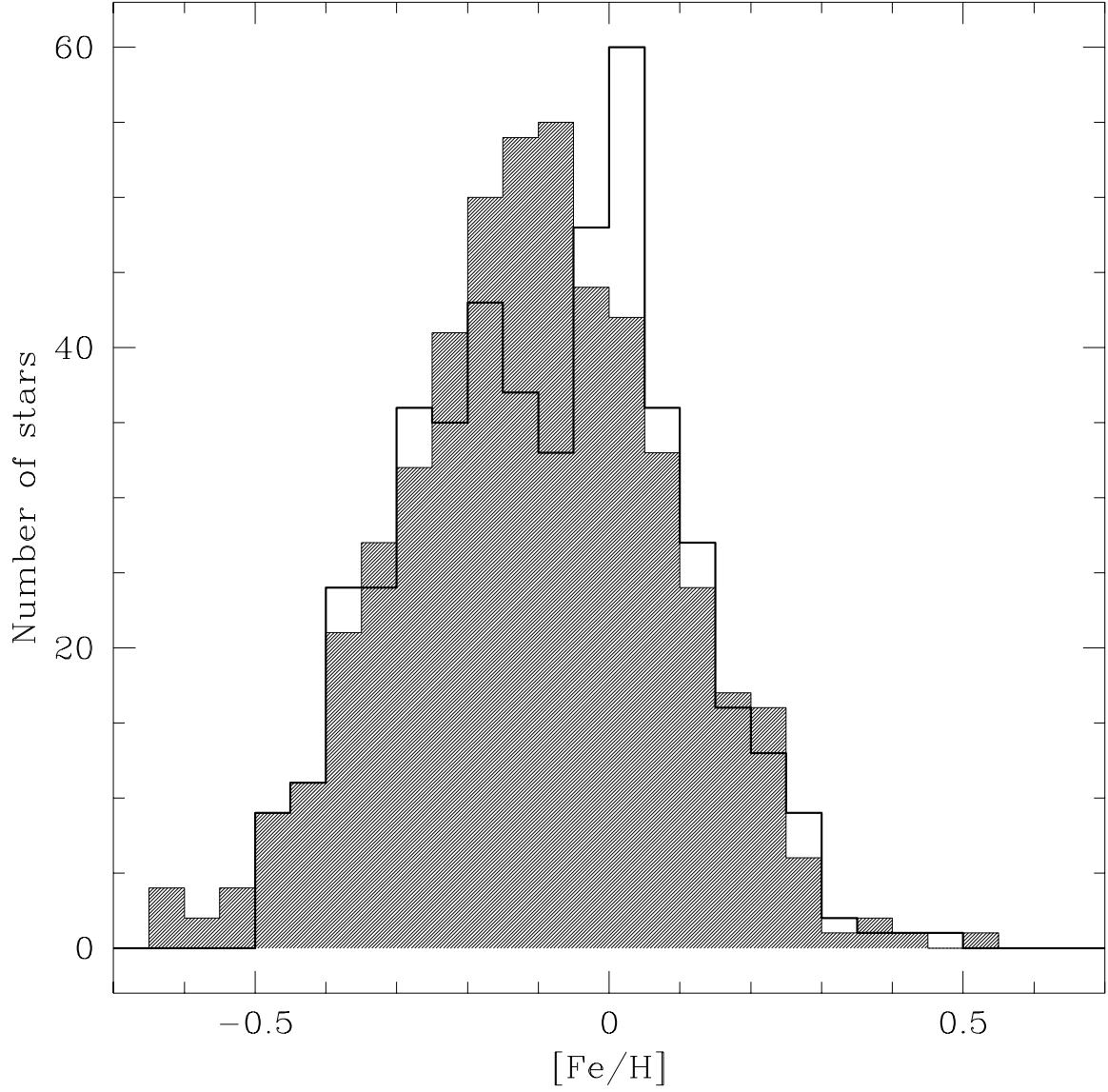}
\caption[Histogram]{The histogram of stellar metallicity for 466
main sequence stars (thick line). The best-fit gaussian (not shown) has mean
$\left<{\rm [Fe/H]}\right>=-0.095$ and variance $\sigma=0.18$. The shaded histogram is
from a Monte Carlo model of 466 stars. These stars are polluted with a
mean mass of $0.4M_\oplus$ of iron. See the text for a more detailed
description of the model. 
\label{Fig_histogram}}
\end{figure}

In Figure~(\ref{Fig_age}) we plot ${\rm [Fe/H]}$ against stellar
age, while Figure~(\ref{Fig_mass}) shows ${\rm [Fe/H]}$ plotted against
stellar mass. The metallicity decreases with increasing age, and
increases with increasing mass. We also note that the Hertzsprung gap
stars tend to have lower metallicity than unevolved stars of the same
age and mass. Figure~(\ref{Fig_histogram}) shows the histogram of [Fe/H]
for our sample. It is well-fit by a gaussian with mean $\left<{\rm [Fe/H]}\right>=-0.095$.

\begin{figure}
\plotone{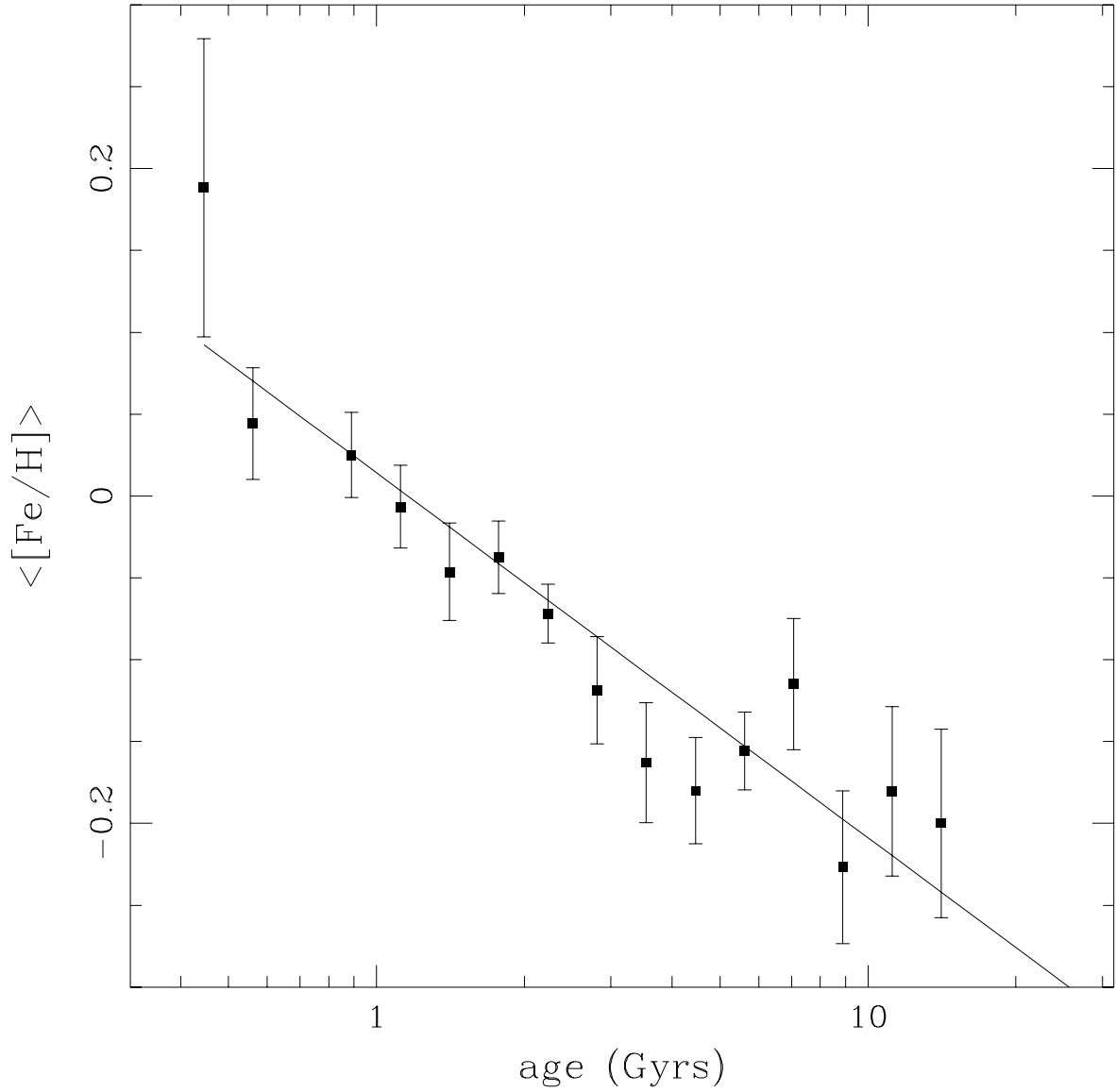}
\caption[Metallicity versus age, binned by age]{Average stellar
metallicity in age bins of width $\Delta\log(age)=0.1$. There are roughly 50
stars per bin. The straight line is a least squares
fit having slope $0.21$ dex/log(Gyr). 
\label{Fig_age_bin}}
\end{figure}

The trend with stellar age is expected on the basis of chemical
evolution; younger stars are constructed using gas that has been
contaminated by material processed in the interiors of massive
stars. The more recent the stellar birth, the higher the fraction of
processed material. The trend is seen even more clearly when we bin
the data by age, as seen in Figure~(\ref{Fig_age_bin}).  The slope of
a least squares fit to the [Fe/H] versus log(age) curve is
$-0.21$. 

\begin{figure}
\plotone{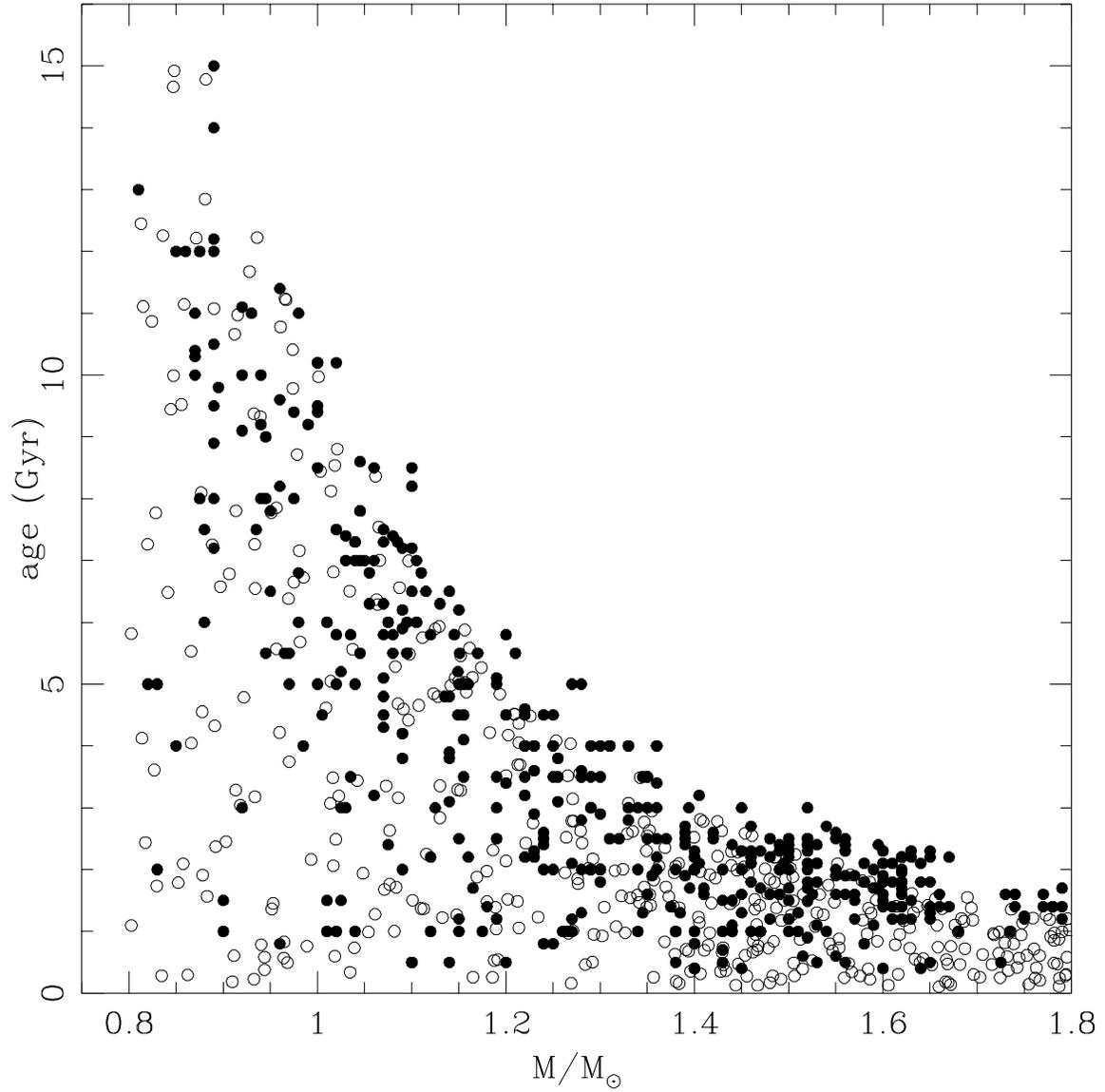}
\caption[Age versus mass]{Stellar age plotted against stellar mass.
Note that on average the more massive stars are younger. Filled
squares are actual stars, open squares are Monte Carlo data. It is
this relation that produces the apparent mass-metallicity trend seen
in the unpolluted Monte Carlo model represented by the dotted line in
Figure \ref{Fig_mass_bin}. 
\label{Fig_age_mass}}
\end{figure}

The trend of increased metallicity with increased stellar mass might
arise from the age-metallicity trend, combined with the short
lifetimes of massive stars. Since the stars we consider are either
unevolved or only slightly evolved, they must have ages less than or
roughly equal to their main sequence lifetimes. Massive stars have
short lifetimes, and hence low ages (see
Figure~\ref{Fig_age_mass}). They will therefore be more metal rich, on
average, than less massive stars.

In the next section we will show that, while part of the variation of
metallicity with stellar mass is due to the age-metallicity relation,
it is not the whole story.

\section{MONTE CARLO MODELS OF POLLUTION\label{gamble}}

The data suggest a possible trend of surface [Fe/H] with stellar
mass. In order to evaluate the significance of this trend, we have
undertaken Monte Carlo experiments to calculate the predicted metallicity
of a population of polluted stars. 

We start by producing a population of unpolluted stars having the same
characteristics as the Cayrel de Strobel-derived sample described
above. This involves generating a sample of stars with mass roughly
uniformly distributed between $0.8$ and $1.8M_\odot$, with
metallicities gaussian distributed about an age dependent mean. From
our sample of stars and using a simple two parameter fit we find that
the mean [Fe/H] is given by
\be 
\left<[Fe/H]\right>=\alpha\times\log(Age)+\beta,
\ee 
where $\alpha=-0.21\pm0.02$ and $b=0.03$. The width of the ${\rm [Fe/H]}$
distribution is slightly age dependent, but we assume an
age-independent width $\sigma=0.18$.

We constrain the stellar age to be less than the calculated lifetime
of the star. As noted above we define the lifetime of a star to be the
age at which the mass of the surface convection zone exceeds ten times
the mass of the mixed surface layer at an age of 100 Myrs.  Our
stellar models have lifetimes (as defined above) which vary as the
$3.65$ power of the stellar mass,
\be 
L(M,{\rm [Fe/H]})=M_1({\rm [Fe/H]})\times \left({M_*\over M_\odot}\right)^{-3.65},
\ee 
where
\be 
M_1({\rm [Fe/H]})=11.5+4.5{\rm [Fe/H]} Gyr
\ee 
is the lifetime of a one solar mass star of metallicity [Fe/H].

Using this prescription we generate 466 stars, yielding the mass-age
relationship depicted by the open circles in Figure \ref{Fig_age_mass};
the agreement with the observed age-mass relationship is very good.

\subsection{The Unpolluted [Fe/H]-Mass Correlation}
We can use this Monte Carlo model to see if the [Fe/H]-mass
correlation seen in our sample could be due simply to the known
age-mass and age-[Fe/H] correlation. To repeat, the idea is as
follows: massive stars are necessarily young. Young stars are metal
rich, since they form out of gas that has been processed through many
generations of high-mass (supernova producing) stars. Thus more
massive stars should have higher metallicity than less massive stars.

This argument is partially borne out by the Monte Carlo model. We have
adjusted the slope of the ${\rm [Fe/H]}$ vs. logarithm of stellar age
relation to obtain agreement with that seen in our sample ($-0.21$ dex
per log(Gyr)), then examined the metallicity as a function of stellar
mass. The stellar metallicities do show a dependence on stellar
mass. However, the slope of the ${\rm [Fe/H]}$ vs mass relation for the
observed stars is $0.26\pm0.03$ dex per solar mass, while that of the
unpolluted Monte Carlo model is $0.18\pm0.03$, different by 3 standard
deviations.

A second indication that there is a real mass dependence in the data
comes from allowing for a third parameter in a least squares
fit. Heretofore we have used a two parameter fit, an age-metallicity
slope and an intercept. If we allow for a linear variation of [Fe/H]
with mass, the slope of the fit is a third parameter. A fit of the
form
\be 
{\rm [Fe/H]}=a_1+a_2\log(age)+a_3\left({M\over M_\odot}\right)
\ee 
does give a significantly improved chi squared.

To see if this three parameter fit can really distinguish between a
pure age dependence and a mixed age and mass dependence, we modified
our Monte Carlo model to allow for a linear dependence of average
metallicity on stellar mass. We then ran the resulting stars through
our least squares routine. If we set the amount of added iron to zero
but force an age dependence, the least squares fit finds the correct
age-metallicity slope. It also correctly finds that there is no mass
dependence, despite the apparent dependence in a metallicity versus
mass plot. If we introduce a mass dependence into the model, the least
squares fit correctly finds both the age and mass
dependences. Finally, we produced samples of stars with a mass
dependence but no age dependence. A plot of metallicity versus age
clearly shows a trend; this is expected since we force the more
massive stars, which are younger, to have higher
metallicities. Despite this apparent age-metallicity trend, the linear
least squares fit finds the correct mass-metallicity slope, and a zero
slope for the age-metallicity relation.

These results suggest that the real data exhibits a mass trend
independent of stellar age, though by itself the argument is not very
convincing. More compelling evidence is obtained by comparing to more
sophisticated models. The inadequacy of a simple age dependent
metallicity is most clearly seen by examining the binned metallicity
versus mass distribution, Figure~(\ref{Fig_mass_bin}). Another
indication is the poor $\chi^2/dof$ of the fit; the best unpolluted
model, shown as the dotted line in the figure, gives $\chi^2=3.2$ for
the binned distribution. We proceed to examine more realistic polluted
models, which give substantially better fits to the data.

\subsection{Realistic Polluted Models}
As noted above, pollution introduces a third free parameter, the
amount of iron added to the star. In the last subsection we assumed
that ${\rm [Fe/H]}$ increases linearly with stellar mass. This is not what
we expect from pollution, since the mass of the surface mixing layer
is not linear, or even monotonic with stellar mass. In this section we
assume a gaussian distribution of accreted mass, with a variance equal
to half the mean. Stars allotted a negative amount of accreted mass by
this process are assumed to have no added material. We assume the
accreted mass is mixed over a surface layer of mass $M_{mix}$, as
described in section \ref{Section_mixing}. Our model for the mass of
the surface mixing layer has many parameters, but we hold them fixed
at the values given in that section. The single adjustable parameter
is then the mean mass of accreted iron. We adjusted this mean to
obtain a minimal $\chi^2$ in the binned data, both metallicity-mass
and metallicity-age. The resulting fit to the metallicity-mass data is
shown as the solid line in Figure (\ref{Fig_mass_bin}). We were able
to find fits with reduced $\chi^2$ equal to one.

The reduction in the $\chi^2$ of the fit afforded by the more
sophisticated pollution model is dramatic. The detailed variation of
${\rm [Fe/H]}$ with stellar mass, including the steep increase starting at
$0.8M_\odot$, the dip around $1.4M_\odot$, and the steep increase to
$1.5M_\odot$ seen in the binned data is reasonably similar to that
seen in the model. We feel that the low $\chi^2$ and the detailed variations
of metallicity with mass offer strong support both for 1) a mass
dependent mixing, as suggested by the Li dip; and 2) for accretion of
$\sim0.5\pm0.2M_\oplus$ of iron in most stars in the solar
neighborhood.

\subsection{Hertzsprung-gap Stars} 
The metallicity-mass trend, combined with the jump in average [Fe/H]
around $1.5$ solar masses strongly suggests that stars in our sample
have accreted iron rich material. However, the properties of our
sample are not well constrained, since the selection criteria of the
various samples collated by Cayrel de Strobel are difficult to
ascertain. Perhaps the apparent variations in average metallicity are
due to some unknown selection effect. For example, one of the
subsamples is that of Favata, Micela, \& Sciortino (1997). A plot of [Fe/H]
versus mass for this sample shows an increase toward large
masses, starting at about $1.15M_\odot$. However, this sample was
selected to have $B-V>0.5$; this eliminates low metallicity high mass
stars, which are bluer than this limit. This is the origin of the
increase in [Fe/H] at $1.15M_\odot$ in this subsample.

We do not see an increase in [Fe/H] at $1.15M_\odot$ in the large
sample. However, perhaps the sharp jump we see at $1.5M_\odot$ is due to
a similar selection effect. The best way to eliminate this
possibility would be to perform an unbiased survey of a large number
of stars in the solar neighborhood. We encourage observers to
undertake such a survey.

We do have available to us a small control sample. Stars that are
slightly evolved, in the sense that their convection zones are 10 or
more times larger than those on the main sequence, but which are not
yet giants (having $M_v$ within $0.5$ of the main sequence value)
provide a check on our interpretation. These stars sit in the
Hertzsprung gap. Unlike giants, their surface gravities and fluxes are
very similar to stars on the main sequence, so their metallicities can
be compared directly to those of main sequence stars.

If pollution is responsible for the [Fe/H] variations over and above
those due to age effects, than Hertzsprung-gap stars should have lower
average metallicities, and they should show no trends with stellar
mass.

\begin{figure}
\plotone{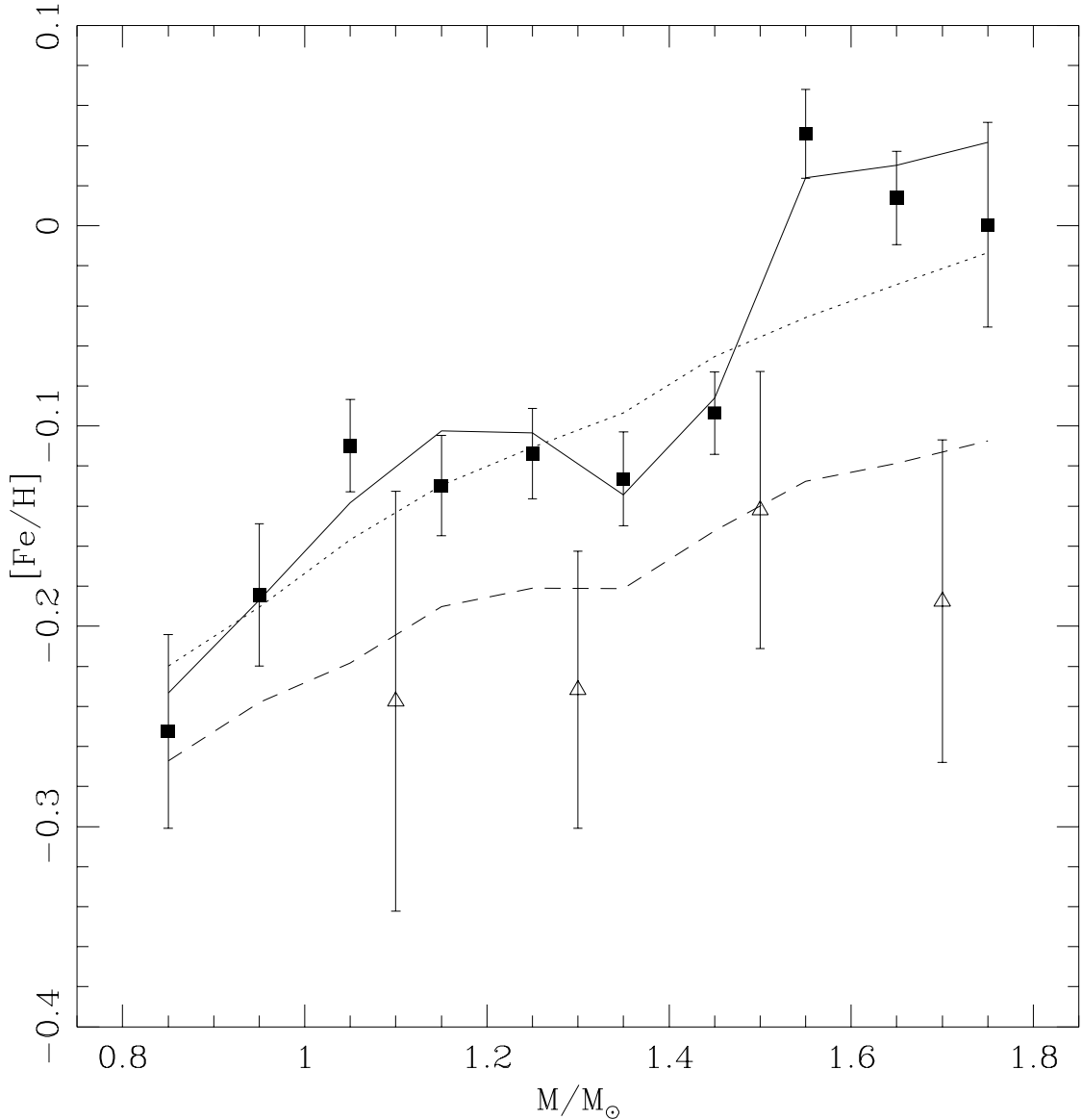}
\caption[Metallicity versus mass, binned by mass]{Average stellar
metallicity in mass bins of width $0.1M_\odot$ (filled squares,
representing unevolved stars). There are roughly 45 such
stars per bin. The open triangles represent the evolved stars (dwarf
stars with surface convection zones ten or more times larger than
$M_{mix}$. There are only 3-6 stars per bin, with a bin width of
$0.20M_\odot$ for these objects. The evolved stars appear to have
lower metallicity than unevolved stars of the same mass.
The dotted line is the unpolluted Monte Carlo model giving the smallest
reduced $\chi^2=3.2$. The solid line is the polluted Monte Carlo model
with an accreted iron mass of $0.4M_\oplus$ and an age-metallicity
slope of $-0.14$; the fit has a reduced $\chi^2=1.0$ The dashed line is
the polluted model after the mass of the surface mixing region has
increased by a factor of ten from its main sequence value, and should
be compared to the evolved star data represented by the open
triangles. This fit also has a reduced $\chi^2$ of order unity.}
\label{Fig_mass_bin}
\end{figure}

The open triangles in Figure~(\ref{Fig_mass}) represent the
metallicities of stars in the gap. It is apparent that these stars
have, on average, slightly lower metallicity than unevolved stars of
similar mass. This is confirmed by an examination of the binned data,
Figure~(\ref{Fig_mass_bin}), which also shows that there is no
evidence for a change in average [Fe/H] in the gap stars at the blue edge
of the lithium gap around 1.5 solar masses. This provides support for
the notion that the variations in metallicity with mass seen in the
unevolved stars is due to pollution.

The dashed line in Fig.~(\ref{Fig_mass_bin}) shows the predicted
${\rm [Fe/H]}$ for the same model as the solid line, but where the mass of
the accreted material has been reduced by a factor of ten. This is
equivalent to increasing the mass of the surface mixing layer by the
same factor. The predicted metallicity for this population is lower
than that predicted for the polluted sample, and shows only a hint of
a ``lithium'' dip. It gives an acceptable fit to the rather sparse
observational data (reduced $\chi^2$  less than one).

\begin{figure}
\plotone{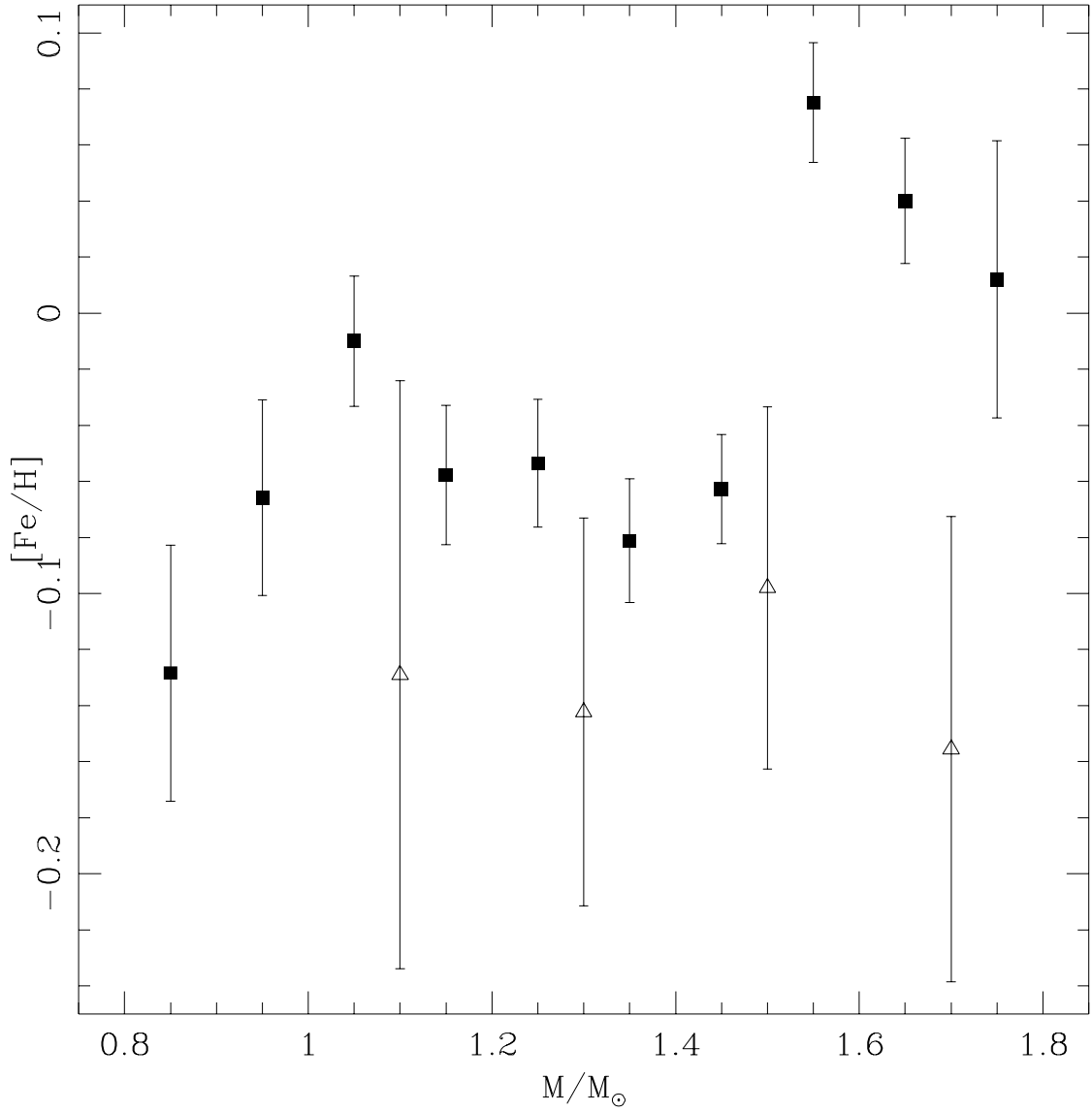}
\caption[Age adjusted data]{ The average metallicity, adjusted for age
using the slope found in our best fit polluted model (a slope of
$0.14$). This figure should be compared to plots of lithium abundance
showing the lithium dip.
\label{Fig_age_Fe_bin_adj}}
\end{figure}

In plotting both the data and the fit we have not adjusted ${\rm [Fe/H]}$
according to the star's age. In Figure~(\ref{Fig_age_Fe_bin_adj}) we
replot the data adjusting for the age-metallicity trend using the
measured slope of the age-metallicity relation in
Figure~(\ref{Fig_age_bin}). This involves increasing the ${\rm [Fe/H]}$ for
old stars. This figure should be compared to plots of lithium
abundance in clusters; we see evidence of a ``lithium'' dip in the iron
data similar to that seen in lithium data. The dip is less distinct in
the iron data, a fact we attribute to the range of metallicities in
our sample (the metallicities in cluster stars show much smaller
dispersions). The variation in metallicity produces a variation in the
mass of the dip, so combining stars with different metallicity will
tend to smear out the dip. This smearing is clearly seen in our Monte
Carlo models (the solid line in Fig.~\ref{Fig_mass_bin}).

We have also examine the metallicities of giants in our sample. They
show a slight increase in [Fe/H] with mass, but this increase is
consistent with the age-[Fe/H] correlation. The average [Fe/H] is
actually slightly {\em larger} than either the unevolved stars or the
Hertzsprung gap stars, although consistent within the rather large
errors. However, given the very different environment for line
formation in these stars, we do not feel it is appropriate to make a
direct comparison between the two populations.

\section{DISCUSSION\label{discussion}}

We made use of three free parameters in fitting the metallicity-age
and metallicity-mass data; one for the slope of the metallicity-age
relation, one (the mean of the added mass) for the correlation in the
metallicity-mass relation, and an overall normalization. We want to
stress that, in fitting the Hertzsprung gap star data, (both age and
metallicity) we did not adjust any free parameters.

We also had a number of parameters in the model which were fixed by
other observations; most were used to describe the location and depth
of the lithium dip. The most critical for our purposes is the mass
$M_{mh}=3\times10^{-3}M_\odot$ of the surface mixing layer in stars
blueward of the dip. Varying this mass will directly affect our
estimates of the amount of iron accreted onto the star. For this
reason our estimate of $0.4M_\odot$ of accreted iron is
uncertain. However, the need to accrete {\em some} material is not
uncertain; reducing $M_{mh}$ will reduce the inferred mass of accreted
material, but models with no pollution will still fail to fit the
data.

The linear model for the mass of the surface mixing layer we employ is
unlikely to be correct in detail; the ``v'' shaped
bottom is particularly questionable. Future work should employ a more
realistic model for $M_{mix}$, but for this first foray we did not
feel that more realistic models were necessary.

If the trends we see are due to pollution, one might surmise that
stars with larger [Fe/H] will form with more massive planetesimal
disks. This might result in larger amounts of iron rich material
falling on stars with larger intrinsic metallicities. This could be
modeled by assuming a correlation between [Fe/H] and the amount of
added material; once again we leave such experiments for later work.

\subsection{Gas Migration and Ingestion of Gas Giant Planets}
The gas giant planets in our solar system appear to have rock/ice
cores with masses of order $10M_\oplus$, with the possible exception
of Jupiter \cite{Guillot_a}, which could lack such a core. These
cores are believed to form at $\gta 5$ AU from the sun since the
surface density of solid matter in the protoplanetary disk is believed
to jump up beyond that distance due to the presence of ice. The
discovery of Jupiter-mass planets in small (less than $0.05$ AU)
orbits around solar type stars motivated a number of groups to suggest
that gas giant planets migrate. If this migration is overly efficient,
Jupiter-mass objects may be accreted onto the star. In the leading
theory of planetary migration, that of tidal interactions between
massive planets and the gas disk out of which they formed, the
migration time is given by the viscous evolution time of the disk,
prompting many authors to predict such accretion \cite{lin,
la}. Others have suggested accretion of iron rich material
originally between the planet and the star \cite{gonzalez97}.

There are two distinct scenarios. In the first, iron rich material is
pushed onto the star by the gaseous disk. This material necessarily
falls on the star before the gas disk disappears, when the star is
between one and ten million years of age. In the second, the enriched
material arrives via some other mechanism, possibly long after the gas
disk disappears. The primary difference between the two, from the
point of view of pollution, is the depth of the stellar convection
zone; stars younger than ten million years and less massive than
$\sim1.3-1.4M_\odot$ have very massive convection zones and
consequently are difficult to pollute.

\begin{figure}
\plotone{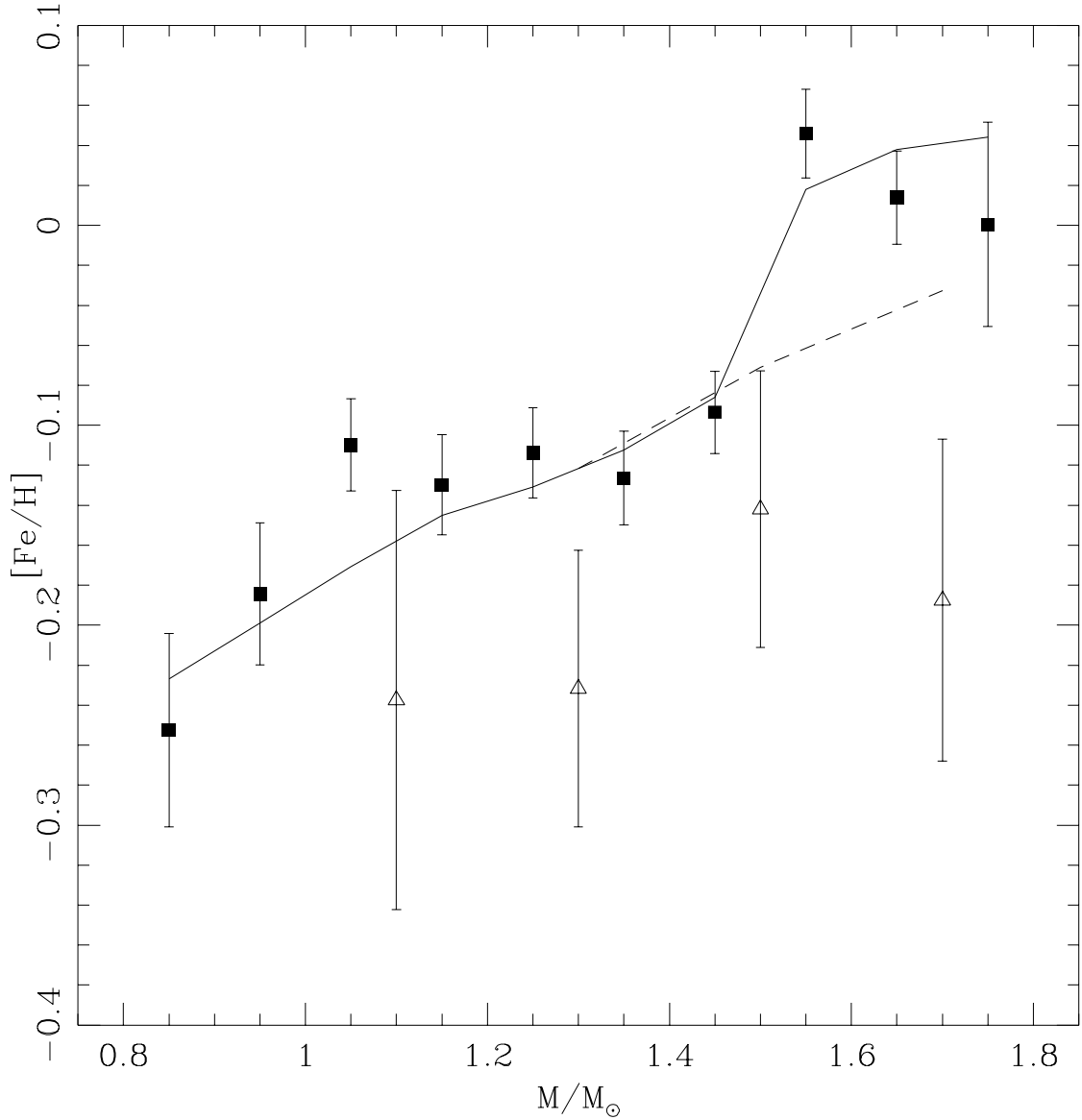}
\caption[Laughlin model]{Metallicity versus mass, showing the same
data as in Figure \ref{Fig_mass_bin} and a model in which the surface
layers of the star are polluted by accretion of enriched material
before the star is 10 million years old. Filled squares are unevolved
stars, open triangles evolved stars. The upper curve is a polluted
model in which a gaussian distributed mass with a mean of
$1.4M_\oplus$ of iron (c.f. the iron content of the giant planets in
table~1) is added to the surface mixing zone of a Monte Carlo
generated sample of stars. For stars lacking convection zones, the
mixing zone is taken to be $3\times10^{-3}M_\odot$. The $\chi^2$ of
the fit is of order unity unity. The lower, dashed, curve gives the
metallicity of the same sample after the mass of the surface mixing
layer has been increased by a factor of ten. This gives a reduced
$\chi^2=2.5$.
\label{Fig_Laughlin}}
\end{figure}

However, as pointed out by Laughlin and Adams (1997), stars more
massive than $1.3-1.4M_\odot$ will show evidence for pollution if they
accrete iron rich material at ages $\sim10$
Myrs. Figure~(\ref{Fig_Laughlin}) shows a Monte Carlo model where
$1.4M_\oplus$ of iron is added to stars before they reach ten million
years of age. We have adjusted the age-metallicity slope and the
fraction of stars accreting material to achieve the best $\chi^2$ fit;
we can find acceptable fits (with $\chi^2\approx 1$) to the unevolved
stellar data as illustrated by the solid line in the figure. Roughly
$1/4$ of these massive stars must accrete Jupiter-analogues as they
approach the main sequence.

This early accretion model does not do so well when compared to the
evolved stars. The model predicts that the metallicities of the
massive stars will lie below those of unevolved stars, as
observed. However the model predicts that low mass stars will not show
any reduction in $\left<{\rm [Fe/H]}\right>$, nor is the drop predicted for the massive
stars as large as that seen in our data. Using our best fit model (the
solid line in Figure~\ref{Fig_Laughlin}) and then increasing the mass
of the surface mixing layer by a factor of ten gives the dashed line
in the figure; for this model the reduced $\chi^2=2.5$; the predicted
metallicities are systematically higher than those observed.

There is also a theoretical difficulty with the early accretion
model. The model is motivated by the possibility that gas in the
protoplanetary accretion disk can push Jupiter-mass objects into the
star. The difficulty is that the gas that does this is also likely to
accrete onto the star. Since any excess iron sequestered in the doomed
planet had to come from the gas in the disk, the gas in the disk must
be iron poor. If this iron poor gas accretes at roughly the same time
as the planet, the net change in the surface metallicity of the star
will be zero.

The fact that the net change in metallicity of the star should be
small, combined with the poor fit of the model to the evolved star
data, suggest that accretion must occur after the accretion disk
vanishes.

\subsection{Ingestion of Gas Giants By Other Means}
Lin (1997) speculates that Jupiter-mass planets might survive the gas
disk only to fall onto the star later. If this occurs, and if these
objects have heavy element abundances similar to gas giants in our
solar system, then the star will gain $\sim1.8M_\oplus$ of iron for
each accreted object. We can find acceptable fits to the data in
Fig.~(\ref{Fig_mass_bin}) if roughly $30\%$ of stars in the solar
neighborhood have each accreted a single Jupiter-mass bodie {\em
after} their convective envelopes thinned. These models naturally
predict that the average metallicity of such stars will decrease by
the observed amount when their convection zones deepen at the end of
their main sequence lifetime.

\begin{figure}
\plotone{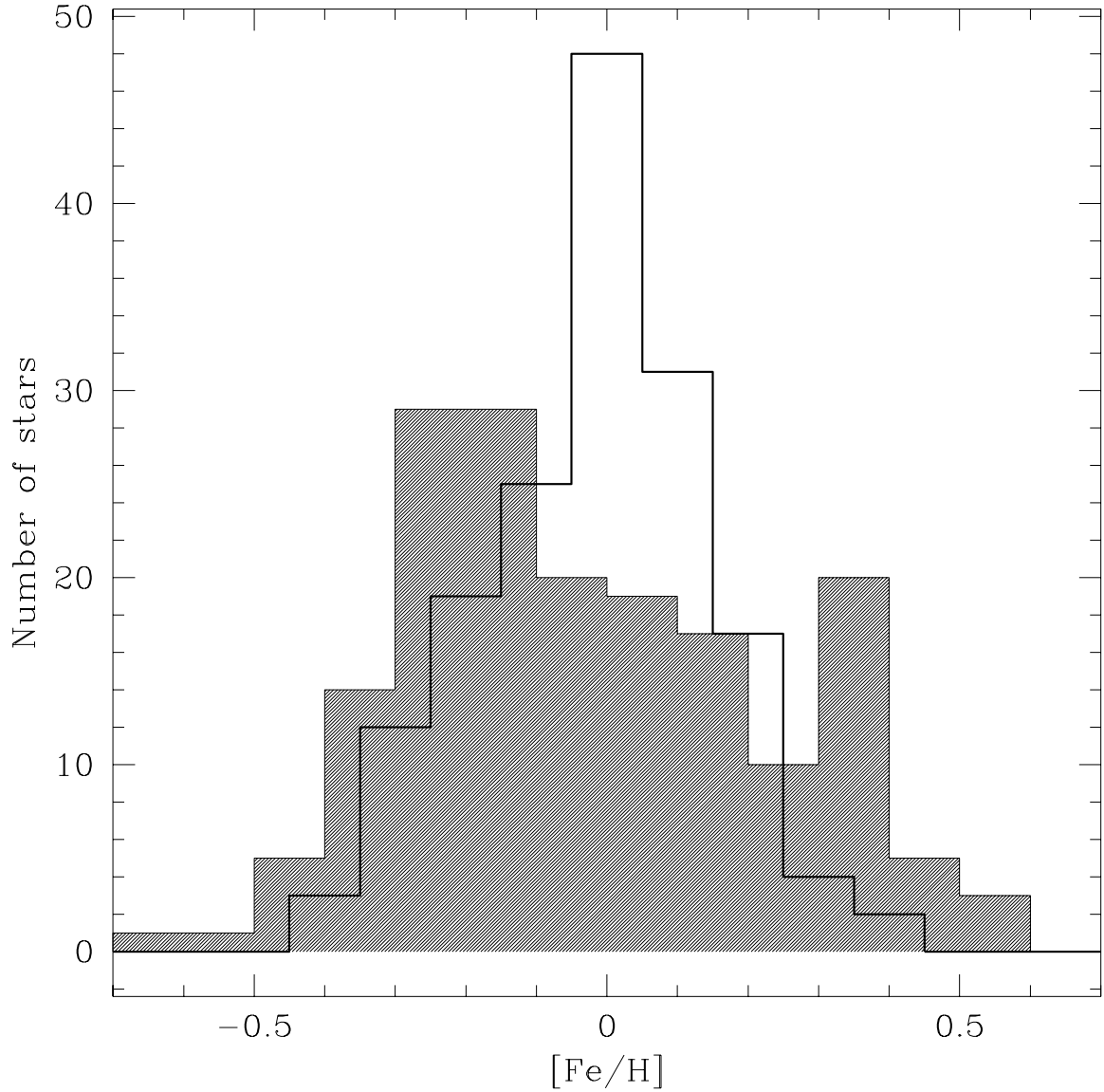}
\caption[]{The histogram of {\rm [Fe/H]} for stars more massive than
$1.4M_\odot$. The thick line is the histogram for the observed stars,
while the light line histogram, which is also shaded, is the result of
a Monte Carlo experiment in which a single Jupiter analogue was
accreted onto $30\%$ of the stars. The bimodal nature of the latter
model arises because we have assumed that the planetary material,
which contains $1.8M_\oplus$ of iron is
mixed with $3\times10^{-3}M_\odot$ of the outer layers of the star;
this results in a large increase in the surface abundance of iron of
those stars that accrete the planets. The mean metallicity of the
entire sample has been forced to match the observed mean
metallicity. The bimodal distribution of the Monte Carlo model is
distinctly different from the observed distribution. 
\label{Fig_histogram_1.4}}
\end{figure}

However we have more information available to us than just the mean
metallicity as a function of stellar mass. In particular, we know that
the distribution of [Fe/H] is well fit by a gaussion, as illustrated
in Figure (\ref{Fig_histogram}). This fact allows us to rule out
pollution dominated by late ingestion of Jupiter-mass planets having
iron contents similar to that of the giant planets in our solar
system. The idea is to use the sensitivity of the high mass stars to
pollution. Adding $\sim1.8M_\oplus$ of iron in discrete lumps to $30\%$
of the stars (to give the observed mean [Fe/H]) will
produce a population  with metallicities about three times
that of the unpolluted population in these high mass stars.
Using the Monte Carlo model that reproduces the variation of
mean metallicity mentioned in the previous paragraph, we obtain the
two-peaked histogram shown in Figure (\ref{Fig_histogram_1.4}). This
can be compared with the observed distribution for those stars above
$1.4M_\odot$ shown on the same figure. If these stars are in fact
polluted, the pollution is distributed smoothly over most of the stars
rather than being concentrated in a fraction of order $30\%$. We
conclude that the metallicity trend we find is not produced by the
late ingestion of Jupiter-mass objects.

\subsection{Surface Metallicity Enhancement By  
Mass Loss or Dust Accretion} 

The solar wind is observed to be iron rich; in fact many elements with
low first-ionization potentials are more abundant, while He, with its
very high first-ionization potential, is under abundant \cite{meyer,
geiss}. If a substantial portion of the stellar convection zone
were lost in a massive wind then the photospheric iron abundance would
be less than the bulk abundance. If less massive stars suffer larger
mass losses (relative to their more massive convection zones) this
would produce a population in which less massive stars have lower
metallicity than more massive stars, as we have found. When these low
mass stars evolve, deep convection zones will form and mix metal rich
material up to the surface; the high mass stars will also show a
slight increase in iron abundance.

However, what we appear to see is that as massive stars evolve, their
metallicities decrease; we don't know what the low mass
stars do when they evolve. There is a second difficulty with this
scenario. The mass loss rates required of low mass stars are
excessive; to lose a substantial fraction of the convection zone
($\sim0.05M_\odot$ for a $0.8M_\odot$ star) the wind must carry away
$\sim5\times10^{-11}M_\odot/yr$ for a billion years. The solar mass
loss rate is $\sim3\times10^{-14}M_\odot/yr$. Recent limits on the
mass loss rates in K dwarfs are much lower, $10^{-12}M_\odot/yr$ or less
\cite{lim, oord}.

Of course the observation that lithium and beryllium depletions are
correlated indicate that mass loss does not extend down to masses of
order $3\times10^{-2}M_\odot$ in stars of mass $\sim1.2-1.4M_\odot$. Taken
together, these facts suggest that mass loss is not responsible for
the metallicity variations we find.

Many young stars of roughly solar or larger mass show
infrared excesses due to circumstellar dust \cite{aumann,
Habing, decin}. Much of this dust may end up on the
star.  Dust particles which absorb photons moving radially and then
radiate them isotropically (in the dust rest frame) lose angular
momentum in a process called the Poynting-Robertson effect. The amount
of angular momentum lost ($\delta \ell$) per photon of frequency $\nu$
is $\delta \ell/\ell \sim h \nu /m c^2$, where $m$ is the rest mass
of the dust particle. Thus, for each accreted dust particle, a star
must emit $\ell/\delta \ell$ photons, which means it must emit a total
energy approximately equivalent to the rest mass energy of the
accreted material. The total energy emitted during the period of dust
accretion is larger than this by $1/\tau$, where $\tau$ is the optical
depth to dust absorption. Thus, for a star radiating at approximately
solar luminosity, the accretion of material will take a time
\begin{equation}
 T_{dust} \sim \frac{10^8 years}{\tau} \left( \frac{M_{acc}}{2
M_{\oplus}} \right) \left( \frac{L_{\odot}}{L} \right)
\end{equation}
For $\tau \sim 1$, this number corresponds roughly to the observed
lifetime of dust around young massive (F) stars \cite{Habing}, but the
spherically averaged $\tau$ is obviously considerably less than 1;
values near $10^{-4}$ are typical. Decin et al. (2000) find longer
lived disks ($\sim1$ Gyr) of similar optical depth, but the maximum
amount of dust that such stars can accrete is still much less than an
Earth mass. We conclude that dust accretion is unlikely to provide
enough material to explain the apparent iron enhancements we see.

\subsection{Accretion of the Interstellar Medium}
Observations of the interstellar medium in the vicinity of the sun
indicate that it has roughly solar abundances ($\sim+0.05$ dex) of
some elements, such as Zn, P, and S \cite{howk}. Iron is typically
depleted, probably onto dust grains. The youngest stars in our sample
have ${\rm [Fe/H]}\approx 0.2$ (Figure \ref{Fig_age}), slightly larger than
the ISM, but our results indicate that about $0.05-0.1$ of this is due
to pollution. This is consistent with the notion that the bulk
metallicity of young stars will equal that of the ISM out of which
they form.

The implication is that older stars have metallicities lower than
that of the ISM in which they currently reside. Accretion of the
relatively metal rich ISM material could in principle produce
metallicity signatures of the type we have found. However, it seems
that the youngest stars in our sample may be metal rich compared to
the present day ISM. In view of the rather large errors in the
metallicity determinations we feel that this conclusion has to
be considered as tentative.

\subsection{$\beta$ Pictoris}
The star $\beta$ Pictoris is surrounded by a dusty disk
\cite{smith}. The dust is believed to have a life time shorter than
the estimated age of the star (between $3\times10^7$ and $10^8$ years)
and so must be replenished, possibly by collisions between
planetesimals \cite{Lecavelier}. Transient red-shifted absorption
features have been seen in spectra of $\beta$ Pic for the last decade
\cite{boggess, lagrange}. These events have been interpreted
as infalling comets or asteroids. Events with very high velocities
($>100$km/s) are seen a few times per month 
\cite{beust, beust_morby}. Similar transients are seen around numerous other
young stars 
\cite{winter, grady}.

The absorption features are believed to be produced by the cometary
tails of the infalling bodies; Beust \& Morbidelli argue that the
infalling bodies must have radii larger than $\sim10$ km in order to
survive the evaporation long enough to evolve onto the inferred highly
eccentric orbits. They estimate that $1.8M_\oplus-18M_\oplus$ of
material is removed from the planetesimal disk by this process if it
has continued for $10^8$ years. They refer to evaporation of these
bodies; however a small but substantial fraction of them will survive
long enough to strike the star. This could provide enough material to
give the signature we have seen.

\subsection{Solar Pollution}
The sun is metal rich for stars of its age and mass, as can be seen
from Figs.~(\ref{Fig_age}) and (\ref{Fig_mass}). The implication is
that the radiative interior of the sun might have a metallicity
slightly lower than the photosphere. A number of authors \cite{Joss,
jcd, levy, jeffery} suggested this possibility
two decades ago as a way out of the solar neutrino problem; lowering
the metallicity of the solar interior reduces the radiative gradient,
which in turn would lower the inferred temperature in the
core. Finally, a lower core temperature would predict that fewer
neutrinos would be emitted compared to the standard solar model.

The metallicity deficit required to explain the lack of solar
neutrinos is much larger than the $0.017$ dex increase resulting from
the accretion of $\sim0.4M_\oplus$ of iron. This avenue for solving
the neutrino problem was dropped by solar workers, partly for this
reason. More recently, \cite{henney_ulrich} examined polluted models
to see if accurate measurements of the the solar five minute
oscillations (p-modes) could be used to detect differences between the
metallicities of the surface convection zone and the radiative
interior. They found that current data was unable to distinguish
between unpolluted models and models with $\sim10M_\oplus$ of accreted
cometary material. The effect on the production of solar neutrinos was
also negligible.

As an aside we note that the SOHO satellite has discovered a large
number of sun grazing stars, a class of comets that plunge into the
sun (see \cite{raymond} and references therein). Roughly 200 have been
observed by SOHO over its lifetime to date. Raymond et al. give an
estimate of about $10^3$cm for the radius of the comet they
observed. If most of the observed objects are of this size, and if we
assume that the currently observed rate has been more or less steady
over the lifetime of the sun, (one strikes the sun roughly every day
as observed in 2000), then the mass of accreted material would be
about one-one millionth of an Earth mass.  Note however that most of
the comets seen by SOHO appear on dynamical grounds to be the result
of the break-up of a single object, so that the currently observed
flux is likely to be a burst of transient activity.

\section{CONCLUSIONS}

We analyzed some 642 stars from the Cayrel de Strobel catalogue having
spectroscopically determined metallicities, and well determined
HIPPARCOS parallaxes. Using a large grid of stellar models, we
determined the age and mass of the stars in our sample. We then
examined the variation of metallicity with stellar age and mass.

We have found striking mass-dependent variations in photospheric iron
abundances of main sequence solar mass ($0.8-1.8M_\odot$) stars in the
solar neighborhood. These variations mimic the variations seen in
lithium abundances. With somewhat less confidence, because of the
small sample size, we find that the iron abundances of Hertzsprung-gap
stars are on average lower than those of main sequence stars, and that
the metallicities of these slightly evolved stars have no mass
dependence. These results are consistent with the accretion of an
average of $\sim0.4M_\oplus$ of iron onto the surface of main sequence
stars. This strongly suggests that terrestrial-type material is common
around solar type stars in the solar neighborhood.

\acknowledgements We would like to thank Barth Netterfield for helpful
conversations. Support for this work was provided by NSERC of
Canada, and by NASA through Hubble Fellowship grant \#HF-01120.01-99A
from the Space Telescope Science Institute, which is operated by the
Association of Universities for Research in Astronomy, Inc., under
NASA contract NAS5-26555. This research made use of the SIMBAD
database, operated at CDS, Strasbourg, France.

\clearpage
\begin{table}
\begin{center}
\begin{tabular}{cccc}
Planet & $M_{Fe}/M_\oplus$ & zone boundaries (AU) & $\Sigma_{Fe}$ ($g\,cm^{-2}$)\\
\tableline
Mercury & $3.3\times10^{-2}$ 	& $0.22-0.56$   &  $1.1$ \\
Venus   & $0.29$                & $0.56-0.86$  	&  $5.7 $ \\
Earth   & $0.38$                & $0.86-1.26$  	&  $4.0 $ \\
Mars    & $0.03$                & $1.26-2.0$ 	&  $0.11$ \\
Asteroids & $0.0005$            & $2.0-3.3$  	&  $6.2\times10^{-4}$ \\
Jupiter & $1.8 $                & $3.3-7.4$	&  $0.36$ \\
Saturn  & $1.7 $                & $7.4-14.4$   	&  $0.11$ \\
Uranus  & $0.9 $                & $14.4-24.7$   &  $0.019$ \\
Neptune & $1.0 $                & $24.7-35.5$   &  $0.013$ \\
\end{tabular}
\end{center}
\caption{Planetary iron content, and the inferred iron surface density
of the minimum mass solar nebula}\label{Table_iron}
\tablecomments{Data for Mercury, Venus, Earth and the asteroids are
from Weidenschilling (1977), for Mars from \cite{longhi}, for Jupiter
and Saturn from \cite{Guillot_b}, for Uranus from \cite{PHS}, for
Neptune from \cite{hpps}. For the gas giants we take representative
values for the heavy element mass; the typical uncertainty is roughly
$\pm30\%$. The zone boundaries are those of Weidenschilling. Using these
values, and assuming solar abundances, we find the iron masses given
in the table.  }
\end{table}

\clearpage

\end{document}